\magnification=1200
\def\Bold{cmbx10 }
\def\Times{cmr10 }

\def\msbm{msbm10 }
\def\msam{msam10 }
\def\cmti{cmti10 }
=\Bold at 14pt
=\Bold at 8pt
=\Times at 8pt
=\Times at 10pt

=\Bold at 12pt
=\Bold at 10pt
=\msbm at 10pt
=\msam at 10pt
\font\eightcal=cmsy10 at 8pt
=\cmti at 10pt
\def\TitleFont{\fourteenbx}
\def\AuthorFont{\tenrm}
\def\AddressFont{\tenrm}
\def\AbstractFont{\eightrm}
\def\AbstractBolsFont{\eightbx}

\def\sec{\sectionfont}        			
\def\ssec{\subsectionfont}          
\def\Bbb{\tenBbb}
\def\LieFont{\tencmti}
\def\ni{\noindent}
\def\ss{\vskip 5pt}
\def\ms{\vskip 10pt}
\def\bs{\vskip 15pt}

\def\nl{\goodbreak\ni}
\long\def\Title#1{\ni{\TitleFont #1}}
\long\def\Abstract#1{\vskip30pt\ni{\leftskip25pt\rightskip10pt{\AbstractBolsFont Abstract:\ }%
{\AbstractFont #1}\par}\vskip20pt}
\long\def\Author#1{\bs\ni{\AuthorFont #1}\par}
\long\def\Address#1{\ss\ni{\AddressFont #1}\par}
%
%
\newcount\PART		          %
\newcount\CHAPTER		          %
\newcount\SECTION		          %
\newcount\SUBSECTION		       %
\newcount\FNUMBER            
\newdimen\TOBOTTOM
\newdimen\LIMIT

\SECTION=0		          
\SUBSECTION=0		       
\FNUMBER=0		          

\def\LastSection{Undefined}
\def\LastSubSection{Undefined}
\def\LastClaim{Undefined}
\def\Last{Undefined}

\def\SectionLabel{\the\SECTION.}
\long\def\NewSection#1{\global\advance\SECTION by 1%
         \bs\ni{\sec  \SectionLabel\ #1}\ss%
         \SUBSECTION=0\FNUMBER=0%
         \gdef\Left{#1}%
         \global\edef\Last{\SectionLabel}%
         \global\edef\LastSection{\SectionLabel}%
         \global\edef\LastSubSection{undefined}%
         \global\edef\LastClaim{undefined}}
\def\SubSectionLabel{\ifnum\SECTION>0 \the\SECTION.\fi\the\SUBSECTION.}
\long\def\NewSubSection#1{\global\advance\SUBSECTION by 1%
         \ms\ni{\ssec #1}\ss%
         \global\edef\Last{\SubSectionLabel}%
         \global\edef\LastSubSection{\SubSectionLabel}}

\def\fopen{(}\def\fclose{)}                               

\def\ClaimLabel{\fopen\ifnum\CHAPTER>0 \the\CHAPTER.\fi%
      \ifnum\SECTION>0 \the\SECTION.\fi%
      \the\FNUMBER\fclose}

\def\NewClaim{\global\advance\FNUMBER by 1%
    \ClaimLabel%
    \global\edef\LastClaim{\ClaimLabel}%
    \global\edef\Last{\ClaimLabel}}


\def\fn{\eqno{\NewClaim}}                         
\def\fl#1{\fn\global\edef#1{\ClaimLabel}}         

\def\cn{\NewClaim}                                
\def\cl#1{\NewClaim\global\edef#1{\ClaimLabel}}    


\def\al{\alpha}

\def\de{\delta}

\def\ep{\epsilon}

\def\te{\theta}
\def\la{\lambda}

\def\om{\omega}
\def\si{\sigma}

\def\De{\Delta}

\def\Te{\Theta}

 \def\R{{\hbox{\Bbb R}}}

 \def\E{{\hbox{\Bbb E}}}

 \def\R{{\hbox{\Bbb R}}}

\def\ip{\hbox to4pt{\leaders\hrule height0.3pt\hfill}
\vbox to8pt{\leaders\vrule width0.3pt\vfill}\kern 2pt}
\def\del{\partial}
\def\arr{\rightarrow}
\def\d{d}
\def\ds{{\bf ds}}
\def\id{\hbox{\rm id}}
\def\dim{\hbox{\rm dim}}
\def\calL{\hbox{$\cal L$}}
\def\scalL{\hbox{\eightcal L}}
\def\calE{\hbox{$\cal E$}}
\def\Lie{\hbox{\LieFont \$}}

\def\then{\Rightarrow}

 \def\,{\mskip\thinmuskip}
 \def\!{\mskip-\thinmuskip}
 \def\>{\mskip\medmuskip}
 \def\;{\mskip\thickmuskip}

%
%
%
%

\def\ni{\noindent}
\def\ss{\vskip 5pt}
\def\ms{\vskip 10pt}

\def\noex{\noexpand}

\def\refs{}
\def\empty{\#}
\def\BibNumber{}
\def\BibTitle{}

\newcount\BNUM
\BNUM=0

\def\bib#1#2{\gdef#1{\global\def\BibNumber{\empty}\global\def\BibTitle{#2}}}

\def\ref#1{#1
\if\BibNumber\empty \global\advance\BNUM 1
\message{reference[\BibNumber]}\message{}
\global\edef\refs{\refs \ss\ni[\the\BNUM]\ \BibTitle}
\global\edef#1{\noex\global\noex\edef\noex\BibNumber{[\the\BNUM]}
 \noex\global\noex\edef\noex\BibTitle{\BibTitle}}
{\bf [\the\BNUM]}
\else
{\bf \BibNumber}
\fi}

\def\Biblio{{\refs}}

\bib{\Ray}{
L.\ Fatibene, M.\ Ferraris, M.\ Francaviglia, R.G.\ McLenaghan
  J. Math. Phys. {\bf 43}(6), 2002, 3147
}

\bib{\Goldschmidt}{
J.\ Rosen,
Annals of Phys. {\bf 69} (1972);
\goodbreak\ni
J.\ Rosen,
Annals of Phys. {\bf 82}(1974); pp.54-69
}

\bib{\Candotti}{
E. Candotti,
Nuovo Cimento v. LXX A, {\bf 2}(1970);
\goodbreak\ni
E. Candotti,
Nuovo Cimento v. 7 A, {\bf 1} (1972)
}

\bib{\FFBRST}{
M.Ferraris, M.Francaviglia, I.Volovich
Il Nuovo Cimento B 110 (3),  (1995)
}

\bib{\Cubo}{
L.\ Castellani, R. D'Auria, P. Fr\'e,
{\it Supergravity and Superstrings. A Geometrical Perspective},
World Scientific, Singapore, 1991
}

\bib{\Saunders}{
D.J. Saunders,$\>$
{\it The Geometry of Jet Bundles},$\>$
Cambridge University Press, Cambridge, 1989 UK
}

\bib{\Marco}{
M.\ Ferraris, Geometrical Methods in Physics, UJEP Brno, Krupka Ed. 1984;
\goodbreak\ni
M.\ Ferraris, M. Francaviglia,
Quaderni di Matematica, Universit\`a di Torino,
Quaderno n.\ {\bf 86}, Torino, 1984;
\goodbreak\ni
	I.Kol{\'a}{\v r},
 Colloquia Mathematica Societatis
	J\'anos Bolyai, 3.1 Differential Geometry, Budapest 1979,
	North-Holland,  (1982);
\goodbreak\ni
	P.L.Garc\'\i a, 
Symposia Math., {\bf 14},
	Academic Press, London,   (1976)
}

\bib{\Libro}{
L. Fatibene, M. Francaviglia,
{\it Natural and Gauge Natural Formalism for Classical Field Theories},
Kluwer (in press)
}

\bib{\Mangiarotti}{
L. Mangiarotti, G. Sardanashvily,\  
{\it Connections in Classical and Quantum Field Theory},
World Scientific, Singapore, 2000
}

\bib{\Olver}{
I.M.\ Anderson, N.\ Kamran, P.J.\ Olver,
  {\it Internal, External, and Generalized Symmetries},
  Advances In Mathematics, {\bf 100}, 1993, 53;
\goodbreak\ni
P.J.\ Olver,
{\it Applications of Lie Groups to Differential
 Equations}, Springer-Verlag New York (1986)
}

\topskip=20pt
\ \vskip 20pt

\Title{On-shell symmetries}

\Author{L.FATIBENE, M.FERRARIS, M.FRANCAVIGLIA}

\Address{Dipartimento di Matematica\nl 
Universit\`a degli Studi di Torino\nl
via Carlo Alberto 10\nl
10123 Torino\nl
ITALY}

\Abstract{
We define on-shell symmetries and characterize them for Lagrangian systems.
The terms appearing in the variation of the Poincar\'e-Cartan form, 
which vanish because of field equations, are found to be strongly constrained 
if the space of solutions has to be preserved.
The behaviour with respect to solution dragging is also investigated in order to discuss relations with the theory of internal symmetries of a PDE. 
}

\NewSection{Introduction}

According to the most general definition, a symmetry of a differential equation is a transformation which preserves the space  of solutions.
If the equation is variational then symmetries (or, more precisely, suitable specific subsets of all symmetries, e.g. Lagrangian symmetries) can be more conveniently discussed in terms of finite dimensional spaces called {\it jet prolongations}.
A satisfactory geometrical framework for Lagrangian symmetries is well established for all
(possibly higher order) field theories (see \ref{\Ray} and references quoted therein).

However, ordinary Lagrangian symmetries are considerably less general than generic symmetries. First of all they are induced by projectable vector fields on the configuration bundle;
moreover they are usually required to leave the Lagrangian (or some Lepagean equivalent object) invariant. 
On the contrary, a generic symmetry is a transformation on the solution space and it is easy to see that there are a number of
such transformations which are not induced by vector fields on the configuration bundle;
moreover, one can easily work out symmetries which preserve field equations without preserving the Lagrangian (e.g. preserving the Lagrangian modulo pure divergences which, as is well-known, do not influence field equations).
As shown in \ref{\Ray}, a reasonable generality in defining {\it generalized Lagrangian symmetries} can be achieved by allowing higher order vector fields which preserve the Poincar\'e-Cartan form modulo contact forms and exact forms.

Nevertheless, other interesting examples can be found outside this last framework. 
To see this, let us consider --as a pedagogical example-- the free particle in one dimension,
described by the Lagrangian
$$
L_{fp}= \hbox{$1\over 2$} v^2
\fn$$ 
over the bundle $\R\times T\R$ endowed with fibered coordinates $(t, q, v)$.
For later convenience we shall also consider higher order tangent bundles, e.g.
$\R\times T^2\R$ endowed with fibered coordinates $(t, q, v, a)$,
$\R\times T^3\R$ endowed with fibered coordinates $(t, q, v, a, b)$ and so on.
Let us consider the following infinitesimal transformation $\de q= \la v + q$ ($\la\not= 0$), which is naturally prolonged to all orders as follows:
$$
\cases{
\de q= \la v + q\cr
\de v= \la a + v\cr
\de a= \la b + a\cr
\dots\cr
}
\fl{\InfinitesimalSymmetry}$$
This transformation preserves the solutions of the free particle Euler-Lagrange equation without preserving the equation of motion itself.
In fact, if one deforms the equation of motion $a=0$ along the transformation $\InfinitesimalSymmetry$ the result $\de a= \la b + a$ is identically vanishing
along solutions.
In \ref{\Ray} it was shown how to regard this infinitesimal transformation as a higher order vector field
$$
\Xi = (\la v + q)\hbox{$\del\>\over\del q$}+(\la a + v)\hbox{$\del\>\over\del v$}
+(\la b + a)\hbox{$\del\>\over\del a$}+\dots
\fl{\SymmetryGenerator}$$
If we consider the deformation of the Lagrangian along the transformation $\InfinitesimalSymmetry$ we obtain
$$
\de L_{fp}= v\de v= v(\la a + v)= v^2+ \la v a
= -aq+\hbox{$\d\over \d t$}\left(vq+\hbox{$\la\over 2$} v^2 \right)
\fl{\FirstSplitting}$$
We stress that there exists an alternative and inequivalent splitting of the Lagrangian variation $\de L_{fp}$ into a total derivative and a term vanishing on-shell. It is simply given by the {\it first variation formula} (see, e.g., \ref{\Ray}), which in this case reads:
$$
\de L_{fp} = v\de v= -a\de q + \hbox{$\d\over \d t$} \left(v\de q\right)=- a(\la v + q)+
\hbox{$\d\over \d t$} \left(\la v^2 + qv\right) 
\fl{\SecondSplitting}$$
This second splitting can be called the {\it trivial splitting}, since such a splitting exists in fact for all Lagrangians and all transformations. Going back to our simple example, notice now that we have been able to show that the Lagrangian remains invariant modulo a total derivative and a term vanishing on-shell in (at least) two different and inequivalent ways.

As was already well-known in the literature (see \ref{\Goldschmidt}, \ref{\Candotti}, \ref{\FFBRST}) any one of these splittings is enough to implement N\"other theorem.
In fact, if the Lagrangian is invariant in the weak sense of
$$
\de L= \al^i E_i + \hbox{$\d f\over \d t$}
\fl{\WeakSplitting}$$
then identity $\WeakSplitting$ can be easily recasted as a conservation law
$$
\hbox{$\d\over \d t$} \left(\hbox{$\del\scalL\over\del u^i$}\de q^i -f\right)= (\al^i-\de q^i)E_i
\fn$$
Reverting then to the free particle, the two splittings above produce by N\"other theorem the following first integrals of motion, respectively:
$$
\calE_{\FirstSplitting}= v(\la v + q)- vq-\hbox{$\la\over 2$} v^2 =\hbox{$\la\over 2$} v^2 
\qquad\qquad
\calE_{\SecondSplitting}\equiv 0
\fn$$ 
The first one being basically the energy of the particle; the second one being trivially {\it ``conserved''} (along any, possibly  non-critical, curve) since it is a constant.
Because of the fact that the trivial splitting actually produces a conservation law of
a very trivial character, it is clear why little effort has been devoted to 
characterize trivial splittings together with their behaviour with respect to solution dragging.

Other more general examples of such a behaviour can be obtained by considering the class of transformations
$$
\de q= B(v)+ A(v)q
\fn$$
where $A(v)$ and $B(v)$ are arbitrary functions.
We also remark that there exist transformations which
also allow a non-trivial splitting as above, although they do not preserve the solution space. As an example, check the following: 
$$
\de q= \la v+ q^2
\fn$$

Passing from Mechanics to Field Theory more physically relevant examples can be found. For instance, supersymmetries in the Rarita-Schwinger model (as well as in other supergravity models; see \ref{\Cubo}) are known to be special kinds of {\it symmetry transformations}, usually called {\it on-shell symmetries}. Remarkably enough, the definition which seems to be implicitly assumed for this notion is the following: {\it a transformation leaving the Lagrangian invariant on-shell modulo pure divergences}.

However, a simple argument shows that this naive attitude is indeed untenable. 
First of all it has to be remarked that {\it all transformations} leave
{\it any} Lagrangian invariant modulo pure divergences and on-shell terms, 
just because of the first variation formula (see equation $(3.2)$ below).

Of course, when a transformation  is claimed in literature to be a symmetry on-shell
some splitting is usually exhibited for the variation of the Lagrangian; however,  its non-triviality is hardly ever proven, while, as we see from the above example, it is clear that  a {\it non-trivial} splitting is strictly speaking necessary.
Nevertheless, we shall see that a non-trivial spitting is not at all {\it sufficient},
since further requirements are needed for a transformation to be a symmetry on-shell.
We shall in fact show that in some cases, even when a non-trivial splitting is exhibited, the transformation might not preserve solutions.

The present paper is therefore devoted to characterize non-trivial on-shell symmetries in Field Theory (in particular in Mechanics) and to provide a geometrical picture able to encompass higher order vector fields as infinitesimal transformations of some kind (we remark that the higher order vector field $\SymmetryGenerator$ does not allow a flow on any $\R\times T^k\R$ nor on the inverse limit $\R\times T^\infty\R$; see below). 
The main result of the present paper is contained in Definition $(3.17)$ which is, to our knowledge, new in the physically oriented literature on the subject.

As a technical tool we shall use jet bundles and Poincar\'e-Cartan forms (see \ref{\Ray}, \ref{\Saunders}, \ref{\Marco},  \ref{\Libro}).
We stress that these are suitable mathematical tools though, in a sense, unessential to our analysis.
Our results can be in fact easily translated back into the usual language of Lagrangian functionals, though loosing some of the geometrical understanding and making some steps considerably more cumbersome.

\NewSection{Notation}

We assume the reader is already familiar with bundle language. Standard references can be found in \ref{\Libro}, \ref{\Mangiarotti} and references quoted therein.
A field theory is defined on a configuration bundle $(B, M, \pi, F)$ with local coordinates $(x^\mu, y^i)$, $\mu=1\dots m=\dim(M)$ and $i=1\dots n=\dim(F)$. 
Configurations are sections $\si:M\arr B$ ($\pi\circ\si=\id_M$).
The bundle of vertical vectors is denoted by $V(\pi)$, $\pi$ being the projection of the relevant bundle. 
The Lie derivative of a section $\si$ with respect to a (higher order) vector field 
$\Xi$ projecting onto an ordinary vector field $\xi$ is defined as
$$
\Lie_\Xi\si= T\si(\xi) - \Xi\circ \si\equiv \left(\Lie_\Xi y^i\right)\del_i
\qquad
\del_i={\del\over \del y^i}
\fn$$

The jet prolongations $J^kB$ take the derivative of fields up to order $k$ into account.
We denote by $\pi^k_h$ the projection of $J^kB$ onto $J^hB$ ($k>h$), by $\pi^k_0$
the projection over $B$ and by $\pi^k= \pi\circ \pi^k_0$ the projection over $M$.
Fibered coordinates on $J^kB$ will be denoted by 
$(x^\mu, y^i, y^i_\mu, y^i_{\mu\nu},\dots, y^i_{\mu_1\dots\mu_k})$, with obvious symmetries in their lower indices.

Being $J^k$ a functor, a strong bundle morphism of $B$ (i.e. a fibered morphism projecting onto a diffeomorfism of $M$) can be canonically prolonged 
to a bundle morphism of $J^kB$. As a consequence vector fields and sections of $B$ can
be canonically prolonged as well to $J^kB$. 
Sections of $J^kB$ which are the prolongation of some section $\si$ of $B$ are called {\it holonomic} and they are denoted by $j^k\si$.

Forms on $J^kB$ which vanish along holonomic sections are called {\it contact forms}.
Contact $1$-forms are generated by
$$
\om^i= \d y^i-y^i_\mu\d x^\mu,\quad
\om^i_\la= \d y^i_\la-y^i_{\la\mu}\d x^\mu,\>
\dots
\fl{\ContactOneForms}$$ 
Contact forms fill a graded ideal in the exterior algebra, where the degree counts
for the number of contact $1$-forms factors $\ContactOneForms$.
For example, $\al_{ij} \>\om^i\land\om^j_\mu\land\d x^\mu$ is a $3$-form of contact order $2$.

A {\it horizontal form} on $J^kB$ is a form which vanishes when contracted along a vertical
vector field. Horizontal forms contain only the differentials of the base coordinates $\d x^\mu$. The coordinate basis of horizontal $m$-forms is denoted by $\ds$ ($m=\dim(M)$);
locally $\ds=\d x^1\land\d x^2\land\dots\land \d x^m$.
By contractions along coordinate vector fields the bases of horizontal $(m-p)$-forms
(with $1\le p\le m$) are generated recursively:
$$
\ds_\mu= \del_\mu\ip \ds,\quad
\ds_{\mu\nu}= \del_\nu\ip \ds_\mu, \quad
\dots
\fn$$
where $\ip$ denotes the interior product between vectors and forms.
By pull-back onto $J^{k+1}B$ any form on $J^kB$ can be canonically split into a horizontal and a contact form
on $J^{k+1}B$.  The projectors along horizontal and contact forms are denoted by $H(\cdot)$ and $K(\cdot)$, respectively.

The jet bundles $J^kB$ form an inverse family and the inverse limit is an infinite dimensional bundle called the {\it infinite jet prolongation bundle}; it is denoted by $J^\infty B$. Accordingly, the projection of $J^\infty B$ over $J^kB$ will be denoted by $\pi^\infty_k$.
This infinite jet bundle is meant to be endowed with the inverse topology and manifold structure. 

A {\it higher order vector field} is a section of the bundle
$(\pi^k_0)^\ast TB\arr J^kB$; see \ref{\Ray}.
Locally a projectable higher order vector field has the following form:
$$
\Xi= \xi^\mu(x^\la)\del_\mu + \xi^i(x^\la,y^i, y^i_\la, \dots, y^i_{\la_1\dots\la_k})\del_i
\fn$$
and the integer $k$ is called the {\it order} of $\Xi$.
Higher order vector fields can be prolonged as ordinary vector fields (see \ref{\Ray}, \ref{\Libro}, \ref{\Olver}).

A {\it Poincar\'e-Cartan form} (PC form) of order $k$ is a form $\Theta$ on $J^{2k-1}B$ obeying the following three axioms:

\itemitem{$PC1$}  $\forall X$, $Y\in V(\pi^{2k-1})$,  $X\ip Y\ip \Theta=0$

\itemitem{$PC2$}  $\forall X\in V(\pi^{2k-1}_{k-1})$,  $X\ip \Theta=0$

\itemitem{$PC3$}  $\forall X\in V(\pi^{2k-1}_{0})$,  $H(X\ip \d \Theta)=0$

\ms
We remark that the space of PC forms is a linear subspace of the
space of all forms on $J^{2k-1}B$. 

As a consequence of the above axioms the horizontal part of a PC form of order $k$ is necessarily of the form:
$$
H(\Theta)= \calL(x^\mu, y^i, y^i_\mu,\dots, y^i_{\mu_1\dots\mu_k} )\>\ds
\fn$$
and it is called {\it the Lagrangian (of order $k$)} induced by $\Te$;
the contact part of $\Te$ has in general a more complicated structure (see \ref{\Marco}, \ref{\Libro} for details).
Often we shall confuse $\Te$ with $(\pi^\infty_{2k-1})^\ast \Te$ which is the expression of the PC form regarded as a form on $J^\infty B$.

If $X=X^i\del_i+X^i_\mu\del_i^\mu+\dots\in V(\pi^{2k-1})$ is a vertical vector then
the quantity 
$$
H(X\ip \d \Theta)=X^iE_i\ds
\qquad\qquad
E_i\equiv \del_i\calL - \d_\mu \del^\mu_i\calL+\dots
\fl{\PCFieldEquations}$$
is directly related to field equations $E_i=0$.
Mechanics can be obtained in the special case $k=1$ and $M\equiv \R$ understanding the
base indices which run on a single value.

From now on we shall explicitly consider the case of a first order field theory ($k=1$),
although our calculation easily extend (with the appropriate modifications) to the higher order case.
In particular, a first order PC form has necessarily the following structure (see \ref{\Ray}, \ref{\Marco}, \ref{\Libro}):
$$
\Te= \calL(x^\la, y^k, y^k_\la)\>\ds 
+ p_i^\mu\> \om^i\land \ds_\mu
\qquad\qquad
p_i^\mu\equiv \del^\mu_i\calL(x^\la, y^k, y^k_\la)
\fn$$
where $\calL(x^\la, y^k, y^k_\la)$ is a first order Lagrangian and $p_i^\mu$ are its {\it canonical momenta}.

\NewSection{Characterization of on-shell symmetries}

Let a {\it N\"other splitting (for a PC form $\Te$)} be a decomposition 
of the Lie derivative of $\Te$  along a higher order vector field $\Xi$ into an exact form $\d\al$,  a contact form $\hat\om$ and an $m$-form $f(E)$ vanishing on-shell, i.e.
$$
\Lie_\Xi\Te= \d \al +\hat\om+ f(E)
\fl{\NoetherSplitting}$$
where $E$ is a short-cut for the Euler-Lagrange equations of the Lagrangian induced by $\Te$.
It is known that these equations take the form $H(X\ip \d\Te)=0$ for arbitrary $X$.
The form $f(E)$ will be assumed to be {\it horizontal} without any loss of generality, since its contact part can be directly included into $\hat\om$.

We stress that the existence of a N\"other splitting is not a restrictive condition on $\Xi$, nor on $\Te$.
In fact, one can trivially show that a N\"other splitting always exists for {\it all} higher order vector fields $\Xi=\xi^\mu\del_\mu+\xi^i\del_i$ and for {\it all} PC forms $\Te$. In fact, one has
$$
\Lie_\Xi\Te= \d( \Xi\ip \Te) + K(\Xi\ip \d\Te)+ H(\Xi\ip \d\Te)
\fn$$
which is a N\"other splitting since $H(\Xi\ip \d\Te)= (\Lie_\Xi y^i)\>E_i\> \ds$ 
does in fact vanish on-shell (being $k=1$, here we have set $\Lie_\Xi y^i\equiv \xi^\mu y^i_\mu-\xi^i$).

 N\"other splittings are introduced in literature since they express a sufficient condition to obtain conservation laws via N\"other theorem (see, e.g., \ref{\Goldschmidt}, \ref{\Candotti}, \ref{\FFBRST}).
We shall investigate whether and when they also preserve solutions.

First of all, let us investigate this matter in a heuristic way.
Let us then consider a $1$-parameter family of morphisms $\Phi_s: J^r B\arr B$ defined
in a neighbourhood of the origin $s\in (-\ep, \ep)$;
furthermore let us assume that for $s=0$ the morphism $\Phi_0$ reduces to the projection $\pi^r_0$.
[The heuristic part of the argument is that, when $r=0$, $\Phi_s$ is the flow of 
an ordinary vector field $\Xi$.
In the event of a higher order vector field the flow (if any exists) would be a flow on the infinite jet prolongation. 
We shall verify below that the results we are going to prove will keep holding true infinitesimally for all higher order vector fields $\Xi$, even when they do not allow a flow on $J^\infty B$].

Let us then drag $\Te$ along the flow $\Phi_s$ (or, better, along its infinite prolongation $J^\infty\Phi_s$). The form so obtained will be denoted by ${}^s\Te\equiv (J^\infty\Phi_s)^\ast \Te$.
We shall here investigate when this flow preserves the solutions (or, better, when it preserves infinitesimally  the solutions).

Our program is achieved in three steps:
\itemitem{(1)} first of all we shall canonically associate a $1$-parameter family of PC forms ${}^s\Te_\ast$ by defining a suitable dragging of $\Te$ along $\Phi_s$ within the space of PC forms;

\itemitem{(2)} we shall determine under which conditions a solution of $\Te$ is a
solution of all the PC forms ${}^s\Te_\ast$ for all $s\in (-\ep, \ep)$;

\itemitem{(3)} we shall verify that the conditions so determined guarantee that 
the infinitesimal generator $\Xi$ (which is a higher order vector field) infinitesimally preserves the space of solutions of $\Te$, even in the event of a higher order vector field which does not allow a flow.

\NewSubSection{Step 1}

The (first order) PC form $\Te$ can be pulled-back along the transformations $J^\infty\Phi_s$ obtaining the forms ${}^s\Te$.
The local expression of a first order transformation is:
$$
\cases{
x^\la= f^\la(x')\cr
y^k= Y^k(x'^\mu, y'^i, y'^i_\la)\cr
y^k_\la= \bar J_\la^\nu\d_\nu Y^k(x'^\mu, y'^i, y'^i_\la)\cr
\dots\cr
}
\qquad\qquad
\cases{
J_\la^\nu= \del_\la f^\nu\cr
J_\la^i= \del_\la Y^i\cr
J_j^i= \del_j Y^i\cr
J_j^{i\la}= \del_j^\la Y^i\cr
}
\fn$$
We shall denote by $\bar J_\la^\nu$ the inverse of $J^\la_\nu$ and 
by $J$ the determinant of $J^\la_\nu$.
One can easily obtain:
$$
{}^s\Te= J\calL\ds' 
+ J \bar J^\nu_\mu p^\mu_j J^j_i\> \om'{}^i\land \ds'{}_\nu
+ J \bar J^\nu_\mu p^\mu_j J^{j\rho}_i\> \om'{}^i_\rho\land \ds'{}_\nu
\fl{\DragPC}$$

We shall first investigate when the form $\DragPC$ is a PC form.
Axioms $PC1$ and $PC2$ are trivially satisfied. Axiom $PC3$ is not satisfied in general,
since for any $X\in V(\pi^\infty_0)$ one has:
$$
H(X\ip \d\>{}^s\Te)= -H(X\ip \d( \De_i^\mu \> \om^i\land \ds_\mu)),\qquad
\De_i^\mu \equiv J E_k J^{k\mu}_i
\fl{\DragEQ}$$
From $\DragEQ$ we see that axiom $PC3$  is satisfied by each ${}^s\Te$ when we restrict to ordinary transformations (i.e., $J^{k\mu}_i=0$).
However, we also see that for a truly higher order transformation there exists a canonical contact form
$$
\De= \De_i^\mu\>\om^i\land \ds_\mu 
\fn$$
such that ${}^s\Te_\ast={}^s\Te + \De$ is a PC form.
In fact, $PC1$ and $PC2$ hold trivially for each ${}^s\Te_\ast$, while $PC3$ follows from $\DragEQ$.
Notice that ${}^s\Te_\ast$ and $^s\Te$ differ by a contact form so that they {\it ``induce''} the same Lagrangian and the same equations.

We shall then define ${}^s\Te_\ast$ to be the dragging (within the space of PC forms) of $\Te$ along $\Phi_s$. The fact that all ${}^s\Te_\ast$ are PC forms will be important for our purposes, since PC forms induce field equations according to equation $\PCFieldEquations$.

\NewSubSection{Step 2}

We have now to implement a condition which ensures the preservation of solutions.
The easiest way to do it, though not the most general, consists in requiring  ${}^s\Te_\ast=\Te$ for all $s$; then the field equations of ${}^s\Te_\ast$ are trivially the same field equations of $\Te$.
Notice however that even when ${}^s\Te_\ast=\Te + \d \al_s$ for some family of $(m-1)$-forms $\al_s$ on $J^\infty B$, 
still they induce the same field equations.
In the Introduction we presented a simple example of a transformation which did not preserve field equations but it still preserved solutions. Generally speaking we have the following Lemma:
 
\ms
{\leftskip 1cm
\ni {\bf Lemma \cn:} let $((a_s)_i^k$, $ (a_s)_i^{k\mu}$, $(a_s)_i^{k\mu\nu}$, $\dots)$ be a $1$-parameter family of (local) functions on $\R\times J^\infty B$; let the limit for $s\arr 0$ be $(\de^i_k,0,0,\dots)$. 

\ni Then $\si$ is a solution for $E_k=0$ iff $\forall s$, $\si$ is a solution of
$(a_s)_i^k E_k + (a_s)_i^{k\mu} \d_\mu E_k+ (a_s)_i^{k\mu\nu} \d_\mu\d_\nu E_k +\dots=0$

\ni\ {\bf Proof:} ($\Rightarrow$) if $E_k=0$ then $\d_\mu E_k=0$, $\d_{\mu\nu}E_k=0$, $\dots$.

\ni\ \phantom{\bf Proof:} ($\Leftarrow$) Set $s=0$.\par
}
\ms

According to this Lemma we say that $\Phi_s$ preserves solutions iff
$$
H(X\ip \d \>{}^s\Te_\ast)\equiv X^i E^{(s)}_i\>\ds =
X^i((a_s){}_i^k E_k + (a_s){}_i^{k\mu} \d_\mu E_k+ (a_s){}_i^{k\mu\nu} \d_\mu\d_\nu E_k +\dots)\>\ds
\fn$$

A necessary and sufficient condition for preserving solutions is hence 
$$
X\ip \d\> {}^s\Te_\ast=
X^i((a_s){}_i^k E_k + (a_s){}_i^{k\mu} \d_\mu E_k+ (a_s){}_i^{k\mu\nu} \d_\mu\d_\nu E_k +\dots)\>\ds + \om^{(1)}
\fl{\StepOneA}$$
for some $m$-form $\om^{(1)}$ of contact order at least $1$ and for all $X\in V(\pi\circ\pi^\infty_0)$.

By simply expanding $X\ip \al= X^iA_i\ds +\om^{(1)}$ for a generic $(m+1)$-form $\al$ and $X\in V(\pi\circ\pi^\infty_0)$,
equation $\StepOneA$ can be recasted as follows:
$$
\d\> {}^s\Te_\ast=
((a_s){}_i^k E_k + (a_s){}_i^{k\mu} \d_\mu E_k+ (a_s){}_i^{k\mu\nu} \d_\mu\d_\nu E_k +\dots)\>\om^i\land\ds + \om^{(2)}
\fl{\StepOneB}$$
for some form $\om^{(2)}$ of contact order at least $2$.

Recalling now that ${}^s\Te_\ast={}^s\Te + \De$ holds, by taking the infinitesimal generator ${\d_s}\big\vert_{s=0}$ and setting $\al{}_i^k=(\dot a_s){}_i^k$, 
$ \al{}_i^{k\mu}=  (\dot a_s){}_i^{k\mu}$, $\dots$, equation $\StepOneB$ is finally recasted as
$$
\d \Lie_\Xi\Te =
(\al{}_i^k E_k + \al{}_i^{k\mu} \d_\mu E_k+ \al{}_i^{k\mu\nu} \d_\mu\d_\nu E_k +\dots)\>\om^i\land\ds - \d (E_k\del ^\mu_i \xi^k \>\om^i\land\ds_\mu)+ \om^{(2)}
\fl{\SymmetryCondition}$$

This condition will be  called {\it covariance identity}. We say that $\Xi$ is a symmetry on-shell if the identity $\SymmetryCondition$ holds true for some choice of the coefficients $\al{}_i^k$, $\al{}_i^{k\mu}$, $\dots$.

We stress that the identity $\SymmetryCondition$ is a strict condition on $\Xi$;
for example, if $\d\Lie_\Xi\Te$ expands with a term $\al_i\>\om^i\land\ds$ which does not vanish on-shell (as it happens for instance with the free particle and the infinitesimal transformation $\de q= \la v + q^2$ mentioned as an example in the Introduction) then the transformation is not a symmetry.

It is now time to show how condition $\SymmetryCondition$ characterizes non-trivial 
N\"other splittings leading to on-shell symmetries.
Let us suppose that $\Lie_\Xi\Te= \d \al +\hat \om + f(E)$.
Because of the inverse limit topology fixed on $J^\infty B$, both $\hat \om$ and $f(E)$ are the pull-back of
objects on some finite order prolongation.
Since we do not need prolongation orders to be sharp, we can assume without any loss of generality that both $\hat \om$ and $f(E)$ truncate at the same order $r$. Hence
$$
\cases{
\hat \om= \te^\la_i\om^i\land \d s_\la + \te^{\la\mu}_i\om^i_{\mu}\land \ds_\la+
\dots + \te^{\la\mu_1\dots\mu_r}_i\om^i_{\mu_1\dots\mu_r}\land \d s_\la + \om^{(2)}\cr
f(E)= (c{}^k E_k + c{}^{k\mu} \d_\mu E_k+\dots
+ c{}^{k\mu_1\dots\mu_r} \d_{\mu_1\dots\mu_s} E_r)\>\ds\equiv C\>\ds\cr
}
\fn$$
Moreover let us set
$$
\al{}_i^k E_k + \al{}_i^{k\mu} \d_\mu E_k +\dots
+ \al{}_i^{k\mu_1\dots\mu_r} \d_{\mu_1\dots\mu_r} E_k\equiv A_i
\fn$$
Hence the covariance condition $\SymmetryCondition$ can be recasted as
$$
\d\hat \om + \d f(E)= A_i \>\om^i\land\ds - \d (E_k\del ^\mu_i \xi^k \>\om^i\land\ds_\mu)+ \om^{(2)}
\fn$$
which in turn expands into the following conditions
$$
\cases{
\del_i C 
	- \d_\la\te^{\la}_i 
	= A_i+ \d_\mu(E_k \del^\mu_i\xi^k)
\cr 
\del^{\mu}_i C 
	- \d_\la\te^{\la\mu}_i 
	- \te^{\mu}_i
	= E_k \del^\mu_i\xi^k
	&$\then  \te^{\mu}_i
		= \del^{\mu}_i C 
		- \d_\la \del^{\la\mu}_i C -E_k \del^\mu_i\xi^k$
\cr 
\del^{\mu\nu}_i C 
	- \d_\la\te^{\la\mu\nu}_i 
	- \te^{\mu\nu}_i
	=0
	&$\then  \te^{\mu\nu}_i
		= \del^{\mu\nu}_i C 
		- \d_\la \del^{\la\mu\nu}_i C$
\cr 
\dots
\cr
\del^{\mu_2\dots\mu_r}_i C 
	- \d_\la\te^{\la\mu_2\dots\mu_r}_i 
	- \te^{\mu_2\dots\mu_r}_i
	=0
	&$\then  \te^{\mu_2\dots\mu_r}_i
		= \del^{\mu_2\dots\mu_r}_i C 
		- \d_\la \del^{\la\mu_2\dots\mu_r}_i C$
\cr 
	  \del^{\mu_1\dots\mu_r}_i C 
	- \te^{\mu_1\dots\mu_r}_i
	=0
	&$\then  \te^{\mu_1\dots\mu_r}_i
		= \del^{\mu_1\dots\mu_r}_i C$
\cr
}
\fl{\SystemA}$$
A solution for $\SystemA$ is obtained by substitution, proceeding from the bottom one up to the top one.
This shows that the contact form $\hat \om$ cannot be fixed at will but it is uniquely determined by $C$ and the symmetry generator $\Xi$.
The first item of $\SystemA$ can be finally recasted as
$$
\E_i(C) = A_i\>,
\qquad\qquad
\E_i\equiv \del_i - \d_\mu\circ \del_i^\mu +\d_{\mu\nu}\circ \del_i^{\mu\nu} + \dots
\fn$$
where $\E_i$ is the Euler-Lagrange operator.
Hence we stress that $\E_i(C)$ is constrained to vanish on-shell and furthermore
to be exactly related to $\d \Lie_\Xi\Te$.

Hence the following definition is well motivated:
\ms
\ni{\bf Definition \cl{\MainDefinition}: }{\it a higher order vector field $\Xi$ (of order $1$) is a symmetry on-shell
of a PC form $\Te$ (of order $1$) if and only if there exists a N\"other splitting
$$
\Lie_\Xi\Te= \d \al + \hat\om + C\>\ds
\fl{\NS}$$
such that $C$ and $\E_i(C)$ both vanish on-shell, $\d \Lie_\Xi\Te= \E_i(C)\>\om^i\land \ds 
- \d (E_k\del ^\mu_i \xi^k \>\om^i\land\ds_\mu)+ \om^{(2)}$
and the contact part $\hat\om$ is uniquely determined as in $\SystemA$.}

\ms
In particular we stress that providing a N\"other splitting $\NS$ and simply checking that $C$ vanishes on-shell (as it is sometimes done in the literature) is definitely insufficient
and sometimes drastically wrong.

\NewSubSection{Step 3}

We shall now verify that the definition of on-shell symmetries given above
holds for higher order vector fields, without resorting to the (possible) existence of their flows.
In particular we shall prove that the variation of field equations along 
$\Xi$ vanishes along solutions.
\ms
\ni{\bf Definition \cn:} We say that a (possibly higher order) vector field $\Xi$ {\it does infinitesimally preserve solutions} of $\Te$ if for all $X\in V(\pi\circ \pi^\infty_0)$
one has
$$
H(\Lie_\Xi (X\ip \d \Te))\simeq 0
\fl{\ipfe}$$
where $\simeq$ means that it does vanish on-shell.
\ms
One can easily expand condition $\ipfe$ into the following
$$
H(\Lie_\Xi (X\ip \d \Te))\simeq
-X^i(\del_k E_i \>\Lie_\Xi y^k +
\del_k^\mu E_i \>\Lie_\Xi y^k_\mu +
\del_k^{\mu\nu} E_i \>\Lie_\Xi y^k_{\mu\nu} 
)\ds
\fn$$
modulo terms vanishing on-shell.
This quantity can be directly shown to vanish on-shell when $\Xi$ is a symmetry on-shell.
In fact, by expanding $\d \Lie_\Xi \Te$ and requiring the on-shell vanishing of the
term along $\om^i\land\ds$ (which is a necessary condition for $\Xi$ to be a symmetry on-shell according to definition $\MainDefinition$) one obtains
$$
\del_k E_i \>\Lie_\Xi y^k +
\del_k^\mu E_i \>\Lie_\Xi y^k_\mu +
\del_k^{\mu\nu} E_i \>\Lie_\Xi y^k_{\mu\nu}\simeq 0
\fn$$ 
i.e. along solutions.

We stress that in definition $\MainDefinition$ as well as in the calculations performed above  the flow of $\Xi$ (which might not exist) is never used.

\NewSection{General vector fields as flow generators}

Higher order vector fields are not vector fields on any finite order prolongation;
however, they prolonge to true vector fields on $J^\infty B$.
In any event, showing the existence of flows on $J^\infty B$ is not trivial 
as in the finite dimensional case and there are explicit examples of vector fields on
$J^\infty B$ which do not allow a flow, even if restricted to a neighbourhood of the parameter origin $s=0$.

Despite for N\"other theorem we need infinitesimal transformations only (as we saw in Section $3$), nevertheless it could be of interest to see whether there exists a group of transformations associated to the infinitesimal
transformation represented by a given higher order vector $\Xi$.

It can be easily shown that any symmetry on-shell is tangent to the infinite prolongation
of field equations $J^\infty E\subset J^\infty B$.
In fact, by expanding $\Xi\ip \d E_i$ one easily obtains
$$
\Xi\ip \d E_i= \del_k E_i \>\Lie_\Xi y^k +
\del_k^\mu E_i \>\Lie_\Xi y^k_\mu +
\del_k^{\mu\nu} E_i \>\Lie_\Xi y^k_{\mu\nu}
\fn$$ 
which, as already proven above, does in fact vanish on-shell.

In Mechanics, when the system is non degenerate the prolongation of the equations of motion is a finite dimensional submanifold of $J^\infty B$.
Moreover, the equations are normal so that they allow one to express any appearence of derivatives higher than one in terms of Lagrangian coordinates and velocities.
As a consequence we obtain a true vector field on the prolongation of the equations, which is
a finite dimensional manifold. Accordingly, it follows that on-shell symmetries define in fact a flow on the equations
(and in turn they can drag solutions) even when they do not define a flow on the whole
prolongation $J^\infty B$.

In Field Theory the situation is a bit more complicated. A vector field is still induced on the submanifold representing the equations, though in general field equations are not normal and the equation submanifold defines an infinite dimensional space.
In field theory, therefore, one is not a priori guaranteed that on-shell symmetries do in fact drag configurations nor solutions.

For instance the higher order vector field related to the transformation $\de q=\la v +q$
prolongs and restricts to $J^\infty E$ to obtain
$$
\bar \Xi=  (\la v +q)\hbox{$\del\over\del q$} +  v\hbox{$\del\over\del v$}
\fn$$
which turns out to be a true vector field on $J^\infty E$ (here parametrized by $(t, q, v)$).
This vector field on $J^\infty E$ defines the following flow
$$
\cases{
t'=t\cr
q'=Q(t,q,v)=(q+\la v s)\>e^s\cr
v'=V(t,q,v)= v \>e^s\cr
}
\fn$$
This can be directly checked to be a $1$-parameter subgroup and to preserve
solutions (i.e. uniform linear motions).

On the contrary, the transformation $\de q= \la v +q^2$ restricts to
$$
\bar \Xi=  (\la v +q^2)\hbox{$\del\over\del q$} +  2vq\hbox{$\del\over\del v$}
+  2v^2\hbox{$\del\over\del a$}
\fn$$
which is not tangent to $J^\infty E$.

\NewSection{Conclusions and perspectives}

We remark that the conditions $\MainDefinition$ above are quite effective in characterizing the non-trivial on-shell symmetries.
For example in the Introduction we presented a number of examples and counterexamples for the free particle. 
For the transformation $\InfinitesimalSymmetry$ we have $\E(-qa)= -2a$ which in fact vanishes on-shell.
The PC form for the free particle is
$$
\Te= \hbox{$1\over 2$} v^2\d t + v\>\om
\fn$$
Hence we obtain
$$
\Lie_\Xi\Te= (v^2+ \la va)\>\d t+  (\la a+2v)\>\om + \la v\>\dot\om =
\d(vq +\hbox{$\la\over 2$}v^2) + (\la a+v)\>\om -q\>\dot\om - aq\>\d t
\fl{\NoethersplittingPf}$$
while the differential is
$$
\d\Lie_\Xi\Te= (\la b+ 2a)\>\d t\land\om + \la a\>\d t\land\dot\om + \om^{(2)}
= -2a\> \om \land\d t  +d(\la a\>\om) + \om^{(2)}
\fn$$
Hence, using the notation of Section $2$, we have $A=-2a$ and $C=-aq$.
Then we can check conditions $\MainDefinition$; we obtain
$$
\E(C)= \E(-aq)= -2a \equiv A
\fn$$
as required.
Furthermore, the contact part $\om$ prescribed by $\SystemA$ is
$$
(\la a+v)\>\om -q\> \dot\om
\fn$$
in complete agreement with what we found in $\NoethersplittingPf$. 
Hence we can conclude that the transformation $\InfinitesimalSymmetry$ is in fact a symmetry on-shell.

As in the Introduction we can also consider the transformation $\de q=\la v + q^2$.
In this case we obtain
$$
\Lie_\Xi \Te=\d(\hbox{$\la\over2$} v^2 + vq^2) + (\la a + 2 q v)\>\om - q^2\>\dot \om - a q^2\>\d t
\fn$$
which we stress is a non-trivial N\"other splitting. Moreover we have
$$
\d\Lie_\Xi \Te= \d(\la a \>\om) - (2 v^2 + 4 qa) \> \om \land \d t + \om^{(2)}
\fn$$
In this second case we hence have $C=- a q^2$ and $A=-2 v^2 - 4 qa$.
Accordingly, we have $\E(C)= -4aq -2v^2 =A$, but $A$ itself does not vanish on-shell as prescribed.
We conclude that the transformation $\de q=\la v + q^2$ is not an on-shell symmetry
even if a non-trivial N\"other splitting exists. 

Further investigation will be devoted to study Rarita-Schwinger supersymmetries.

\NewSection{Acknowledgements}

We are grateful to Prof. A. Borowiec (Wroc\l aw-Poland), Prof. L. Lusanna(Arcetri-Italy) and Prof. R. Mann (Waterloo-Ontario) for their interesting and useful comments.

\NewSection{References}

\Biblio

\end